\documentclass[%
 amsmath,amssymb,
 reprint,%
]{revtex4-1}

\usepackage[utf8]{inputenc}
\usepackage{amsmath,amssymb}
\usepackage{graphicx}
\usepackage{xcolor}

\usepackage{changes}
    \colorlet{Changes@Color}{blue}

\begin{document}

\title{Trion induced photoluminescence of a doped MoS$_2$ monolayer}

\author{Yaroslav V. Zhumagulov}
\affiliation{ITMO University, St. Petersburg, 197101, Russia}
\affiliation{University of Regensburg, Regensburg, 93040, Germany}

\author{Alexei Vagov}
\affiliation{Institute for Theoretical Physics III, University of Bayreuth, Bayreuth 95440, Germany}
\affiliation{ITMO University, St. Petersburg, 197101, Russia}

\author{Paulo E. Faria Junior}
\affiliation{University of Regensburg, Regensburg, 93040, Germany}

\author{Dmitry R. Gulevich}
\affiliation{ITMO University, St. Petersburg, 197101, Russia}

\author{Vasili Perebeinos}
\email{vasilipe@buffalo.edu}
\affiliation{Department of Electrical Engineering, University at Buffalo, The State University of New York, Buffalo, NY 14260, USA}
\affiliation{ITMO University, St. Petersburg, 197101, Russia}

\begin{abstract}
We demonstrate that the temperature and doping dependencies of the photoluminescence (PL) spectra of a doped MoS$_2$ monolayer have several peculiar characteristics defined by trion radiative decay.  While only zero-momentum exciton states are coupled to light, radiative recombination of non-zero momentum trions is also allowed. This leads to an asymmetric broadening of the trion spectral peak and redshift of the emitted light with increasing temperature. The lowest energy trion state is dark, which is manifested by the sharply non-monotonic temperature dependence of the PL intensity. Our calculations combine the Dirac model for the single-particle states, the parameters for which are obtained from the first principle calculations, and the direct solution of the three-particle problem within the Tamm-Dancoff approximation.  The numerical results are well captured by a simple model that yields analytical expressions for the temperature dependencies of the PL spectra.{}
\end{abstract}

\maketitle

\section{Introduction}

Monolayers of transition metal dichalcogenides have attracted a lot of attention due to their unique electronic and optical properties, which differ significantly from those in bulk materials. A representative example is a monolayer of MoS$_2$.~\cite{Mak2010, Splendiani2010}  This structure is a non-centrosymmetric 2D semiconductor that has two degenerate direct gaps at both non-equivalent $K$ points in the Brillouin zone.~\cite{Ramasubramaniam2012,Xiao2012,Kormnyos2015} Large spin-orbit splitting in the valence and conduction bands in the vicinity of the band edges allows efficient control of the spin and valley degrees of freedom.~\cite{Mak2012nano,Zeng2012,Cao2012,Sallen2012} Strong light-matter coupling and other properties can be exploited in a range of optoelectronic devices, including phototransistors,~\cite{LopezSanchez2013} logic circuits,~\cite{Radisavljevic2011, Wang2012} light-emitting and -harvesting devices.~\cite{Feng2012, Cheng2014, LopezSanchez2014, Pospischil2014, Ross2014}

Spectral characteristics of MoS$_2$ monolayers reveal an increased role of many-particle states such as  excitons~\cite{Mak2010} and charged trions.~\cite{Mak2012} The latter can be controlled by applying an external electric field.  The 2D dimensionality enhances the Coulomb interaction compared to conventional bulk semiconductors such as GaAs,~\cite{Yu2010} giving rise to much larger exciton~\cite{Mak2010,Splendiani2010,Komsa2012,Feng2012,Qiu2013,BenAmara2016} and trion binding energies.~\cite{Mak2012,Ross2014,Lui2014,Rezk2016,Zhang2014,Mouri2013,Singh2016,Scheuschner2014,Soklaski2014,Zhang2015_2}  Signatures of negatively charged trions were observed in the photoluminescence (PL) spectrum of a doped MoS$_2$ monolayer.~\cite{Mak2012} However, the analysis of experimental data is hindered by the complexity of the system, in particular, of its single-particle band structure which gives rise to a rich variety of possible many-particle (exciton and trion) states. An adequate interpretation of the experiments demands a detailed theoretical investigation of exciton and trion excitations using realistic models for the electronic bandstructure.~\cite{Wang2018}

The two-particle exciton and three-particle trion states are fundamentally different. Due to the momentum conservation law only excitons with zero (or very close to zero) center-of-mass momentum can be created by external light or recombine radiatively.  However,  this restriction does not apply to trions because the final single-particle state can carry a non-zero momentum. This results in a qualitative difference between the exciton- and trion-induced branches of the optical PL spectrum, that can be detected experimentally.

The goal of this work is to quantify the role of trion radiative recombinations in a doped MoS$_2$ monolayer on the lower energy domain of the PL spectrum. Of particular interest are the temperature and the doping dependencies of the trion peak characteristics such as transition energy, intensity, shape, and width. The calculations of the trion states are done employing an effective Dirac model for the single-particle states with parameters fitted to the {\it ab initio} calculations for the MoS$_2$ bandstructure. The results are interpreted within a simple phenomenological model for the trion states with parameters extracted from the numerical calculations.

\section{Method and model}

The calculation of trion states is a non-trivial task even in 2D,  especially in the presence of doping which makes the problem a truly many-body one.  Since the first observation of trions in MoS$_2$ monolayers, various approximate theory approaches have been employed to describe them, including stochastic and variational methods,~\cite{Berkelbach2013,Kezerashvili2016,Kidd2016,Zhang2015,VanTuan2018} Monte Carlo,~\cite{Mayers2015} path integral,~\cite{Kylnp2015,Velizhanin2015} and diagrammatic expansions.~\cite{Efimkin2017,Sidler2016,Back2017,Scharf_2019} Due to rapid advances of the computational methods and computing power direct diagonalization of the corresponding three-particle Hamiltonian is now possible.~\cite{Deilmann2016,Drppel2017,Deilmann2017,Deilmann2018,Deilmann2018BP,Arora2019,Torche2019,Tempelaar2019,Zhumagulov2020} However, until now the main attention has been focused on trions with zero center-of-mass momentum. Certain progress in the studies of non-zero momentum neutral (two-particle) excitons was achieved for undoped systems.~\cite{Yu2014,Qiu2015, Cudazzo2016, Steinhoff2017} In this work we take into account trions with non-zero momentum by extending the approach developed earlier to study trions in doped transition metal dichalcogenide monolayers.~\cite{Zhumagulov2020,Tempelaar2019}

\subsection{Single-particle states}

Single-particle states of the MoS$_2$ structure, that provide a basis for the subsequent solution of the three-body problem for trions, can be obtained by standard {\it ab initio} calculations using the DFT-GW approaches. However, the analysis of the low-energy trion states~\cite{Drppel2017,Zhumagulov2020} demonstrates that they comprise only the single-particle states in the vicinity of the two points of the Brillouin zone, $K$ and $K^\prime = -K$, while the contribution of all other states is negligibly small. Thus, the calculations of trions can thus be simplified considerably by assuming an effective single-particle massive Dirac model,~\cite{Xiao2012} well suited to describe the single-particle band structure in the vicinity of those two points.  Despite its relative simplicity, the model captures all relevant phenomena such as the coupling between the spin and the valley degrees of freedom.

The massive Dirac Hamiltonian, that describes single-particle states in the valleys $K$ and $K^\prime$, includes the Zeeman-type spin-orbit coupling (SOC) and reads as~\cite{Xiao2012}
\begin{align}
    H_0&=g \big( \tau k_x\sigma_x+k_y\sigma_y \big) + \frac{\Delta}{2} s_0\otimes\sigma_z \notag \\ 
    &+ \tau s_{z} \otimes \big( \lambda_c \sigma_{+} + \lambda_v \sigma_{-} \big),
\label{HDirac}
\end{align}
where $g = a t_h$ is the effective coupling constant, $a$ is the lattice constant, $t_h$ is the effective hopping, $\Delta$ is the bandgap, $\sigma$ are pseudospin Pauli matrices acting in the band subspace, $s_z$ is the spin Pauli matrix acting in the spin subspace, with $s_0$ denoting the unity matrix. The Hamiltonian (\ref{HDirac}) is written in the basis  $\big\{ \big| \Psi_{c \uparrow}^\tau \big\rangle, \big| \Psi_{v\uparrow}^{\tau} \big \rangle, \big| \Psi_{c\downarrow}^\tau \big \rangle, \big | \Psi_{v \downarrow}^\tau \big \rangle \big\}$ which corresponds to the states in the conduction band $c$ and valence band $v$ with spins $\uparrow \downarrow$ at $K$ $(\tau=1)$ and $K^{\prime}$ $(\tau=-1)$ valleys. The last contribution in Eq. (\ref{HDirac}) describes the SOC. 

Parameters of the Hamiltionian (\ref{HDirac}) are extracted by fitting its spectrum with the results of the {\it ab initio}  calculations for the band structure of a stand-alone MoS$_2$ monolayer,~\cite{Berkelbach2013,Cho2018,Zollner2019} where we assume the lattice constant $a=3.19$\ \AA. Results of these calculations yield $t_{h}=1.41$ eV, $\Delta =2.67$ eV, $\lambda_c=1.5$ meV and $\lambda_v = 73$ meV.


\subsection{Trion states}

Quantum states $t$ of negatively charged trions are obtained by solving the eigenvalue problem for the three-particle Hamiltonian derived by spanning the full many-body Hamiltonian onto the space of trion states constructed as linear superpositions
\begin{align}
  \left| t \right\rangle = \sum_{c_1,c_2,v} A_{c_1 c_2 v}^{t}  \left| c_1 c_2 v \right\rangle, \,  \left| c_1 c_2 v \right\rangle =  a_{c_1}^\dagger  a_{c_2}^\dagger a_{v}^\dagger \left| 0 \right\rangle,
  \label{eq:expansion}
\end{align}
where $c_{1,2}$ denote electron states in the conduction band, $v$ are hole states in the valence band and the double counting is avoided by imposing the restriction $c_1<c_2$. The corresponding three-particle wavefunction is constructed from the single-particle functions $\phi_{c,v}(x)$ as
\begin{align}
    \Psi^{t}(&x_1,x_2,x_3)=\frac{1}{\sqrt{2}} \sum_{c_1,c_2,v} A_{c_1 c_2 v}^{t} \phi_{v}^{*}(x_3) \notag \\
    &\times \big[ \phi_{c_1}(x_1)\phi_{c_2}(x_2)-\phi_{c_2}(x_1)\phi_{c_1}(x_2) \big].
\label{eq:wave_function}
\end{align} 
Coefficients $A_{c_1 c_2 v}^{t}$ in the expansion are found by solving the  matrix eigenvalues problem
\begin{align}
\sum_{c_1^\prime c_2^\prime v^\prime} {  H}_{c_1c_2v}^{c_1^\prime c_2^\prime v^\prime} A_{c_1^\prime c_2^\prime v^\prime}^{t} = E_t A_{c_1c_2 v}^{t}.
\label{eq:hamiltonian}
\end{align}
The Hamiltonian matrix has three contributions ${  H} = {  H}_{0} + {  H}_{cc}  + {  H}_{cv}$ defined as
\begin{align}
\label{eq:three_particle}
{  H}_{0}=&(\varepsilon_{c_1} + \varepsilon_{c_2} -\varepsilon_{v}) \delta_{c_1c_1^\prime} \delta_{c_2c_2^\prime}  \delta_{v v^\prime}, \\
{  H}_{cc} =&(W_{c_1c_2}^{c_1'c_2'}-W_{c_1c_2}^{c_2'c_1'})\delta_{vv'}, \nonumber \\
{  H}_{cv}=&-(W_{v'c_1}^{vc_1'}-V_{v'c_1}^{c_1'v})\delta_{c_2c_2'}-(W_{v'c_2}^{vc_2'}-V_{v'c_2}^{c_2'v})\delta_{c_1c_1'} \nonumber \\
& +(W_{v'c_1}^{vc_2'}-V_{v'c_1}^{c_2'v})\delta_{c_2c_1'}+(W_{v'c_2}^{vc_1'}-V_{v'c_2}^{c_1'v})\delta_{c_1c_2'}, \nonumber
\end{align}
where $\varepsilon_{c,v}$ denotes the single-particle energy of the particle/hole states, $W$ and $V$ are the screened and bare Coulomb matrix elements, respectively. This approach is a direct extension of the Tamm-Dancoff approximation for two-particle excitons onto three-particle trions. Matrix elements for the bare Coulomb interaction are given by
\begin{equation}
\label{eq:Coulomb}
    V^{ab}_{cd} = V(\textbf{k}_a -\textbf{k}_c) \langle u_c | u_a \rangle \langle u_d | u_b \rangle,
\end{equation}
where $V ({\bf q}) = 2\pi e^2/ \varepsilon q$ is the Fourier component of the bare Coulomb potential and $\langle u_c| u_a \rangle$ is the overlap of the single-particle Bloch states $c$ and $a$. The screened potential is given also by Eq. (\ref{eq:Coulomb}), however, instead of the bare Coulomb potential $V({\bf q})$ one substitutes the standard Rytova-Keldysh  expression\cite{rytova1967the8248,keldysh,Cudazzo2011}
\begin{align}
\label{eq:Coulomb_screened}
   W({\bf q}) =  \frac{2\pi e^2}{\varepsilon q(1+r_0 q)},
\end{align}
where $r_0$ is the screening length. In the calculations we assume the effective dielectric constant is $\varepsilon =1$ (stand-alone monolayer in vacuum),  and $r_0 =42$ \ \AA.~\cite{Berkelbach2013,Cho2018}

The system has several conserving quantities. One of which is the total moment of a trion 
\begin{equation}
   {\bf k} = {\bf k}_{c_1} + {\bf k}_{c_2} - {\bf k}_v.
\end{equation}
Another one is a total spin and its $z$ axis projection
\begin{equation}
    S_{z} = s_{z}^{c_1} + s_{z}^{c_2} - s_{z}^v.
\end{equation}
These can be used to reduce the computational efforts by taking into account only three-particle states with the fixed total momentum and the $z$-component of the total spin. 

Furthermore, the calculations can be restricted to trions with a total spin $1/2$, as only such states allow for the radiative recombination of electron-hole pairs (optically active bright states). We neglect scattering between the states with the total spin $1/2$ and $3/2$, which could be facilitated by magnetic impurities.  

Following the conventional classification, one distinguishes two types of trion states - with $|s_{z}^{c_1} + s_{z}^{c_2}|=1$ and $0$, respectively.~\cite{Yu2014,Mayers2015,Courtade2017,Torche2019,Tempelaar2019} One can also define the intra- and inter-valley states as those with  $|\tau_{c_1} + \tau_{c_2}|=2$ and $0$~\cite{Torche2019,Tempelaar2019}. The fermionic trion states can also be classified by the product $S_z  \tau$ of the total spin $S_z$ and the total valley $\tau = \tau_{c_1} + \tau_{c_2} - \tau_v$  quantum numbers. A necessary condition for a state to be bright is $S_{z} \tau =   \pm 1/2$. However, not all such states can decay radiatively.

Description of trion and exciton states in a doped monolayer requires solution of the truly many-body problem, where the trion is affected by the doped electrons. Due to the enormous complexity of this problem, its complete solution is not feasible. However, we employ the approach which circumvents this difficulty by solving the eigenvalue problem for three particles on a uniformly discretized mesh in the Brillouin zone.~\cite{Zhumagulov2020} Within this approach the doping density  $n=g_v g_s/(\Omega_0 N^2)$ is related to the number of mesh points $N\times N$ and the area of the primitive cell $\Omega_0$ ($g_s$ and $g_v$  are the spin and valley degeneracies, respectively). The corresponding Fermi energy of the doped electrons $E_F=\hbar^2\Delta k^2/2m$, where the discretization interval for the hexagonal lattice of MoS$_2$ is $\Delta k = 4\pi/(\sqrt{3}aN)$ with the effective electron mass $m=0.5$ $m_e$ used here.  

\begin{figure}
\begin{center}
\includegraphics[width=0.4\textwidth]{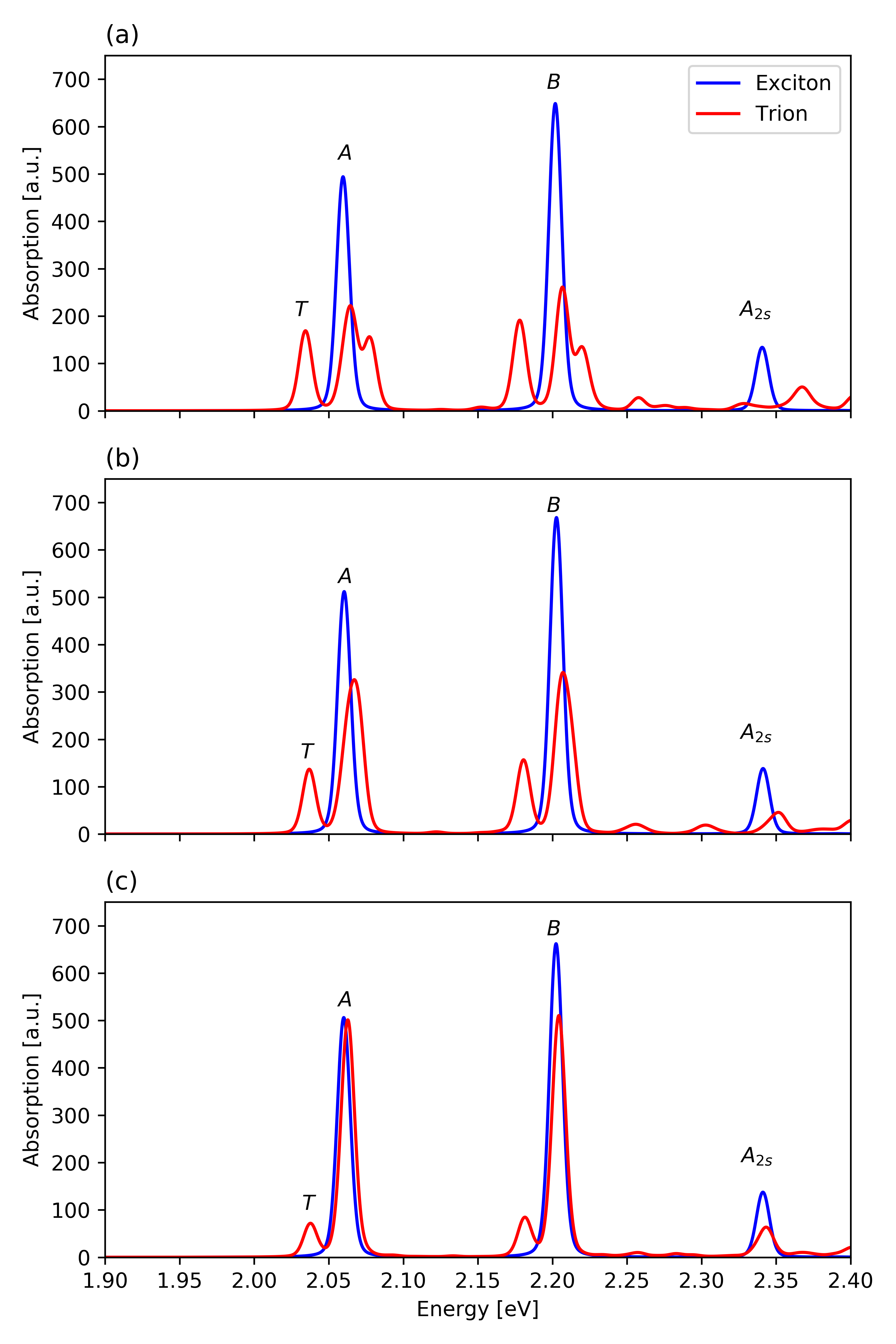}
\end{center}
\caption{Linear absorption spectrum of a doped MoS$_2$ monolayer (red line), calculated for different doping levels: (a) $n=19.7 \times 10^{11}$ cm$^{-2}$  ($E_F=17.1$ meV or $48 \times 48$ k-mesh), (b) $n=12.58 \times 10^{11}$ cm$^{-2}$ ($E_F=10.9$ meV or $60 \times 60$ k-mesh) and (c) $n=5.6 \times 10^{11}$ cm$^{-2}$ ($E_F=4.9$ meV or $90 \times 90$ k-mesh). The lowest T peak is due to trion states. For comparison the blue lines show the spectra of an undoped MoS$_2$ monolayer with the peaks A, B and A$_{2s}$ due to exciton states.  }
\label{fig:abs}
\end{figure}

\subsection{PL spectrum}

In the calculations of the PL spectrum, we assume that the monolayer is excited off-resonantly, i.e. the photon energy is well above the lowest exciton or trion transition energies. The non-equilibrium state, created by the excitation,  undergoes a fast relaxation towards the quasi-equilibrium thermal distribution. Then the trions start to recombine much slower radiatively. 

Within this picture, the PL spectrum is calculated using standard Fermi's rule for the transition rates assuming the Boltzmann distribution for trions in the quasi-equilibrium state. This yields the following expression for the PL spectrum:
\begin{equation}
    L\left(\hbar\omega\right)\propto\frac{1}{Z}\sum_{\bf k}  \sum_{t,c} e^{- \frac{E_{t}({\bf k})}{k_B T}} 
    \big \vert {\bf P}_{t}^{c}({\bf k}) \big \vert^2 \delta \big( \hbar \omega - \Delta_{ t}^{c} ({\bf  k}) \big),
    \label{eq:PL_spectrum}
\end{equation}
where $T$ is a temperature of the quasi-equilibrium distribution. Dipole matrix elements  ${\bf P}_{t}^{c}  =  \langle  c \vert \textbf{p} \vert t \rangle$ in this expression are calculated for trion $t$ and single electron $c$ states with the same momentum ${\bf k}$, such that: 
\begin{equation}
    {\bf P}_{t}^{c}({\bf k})  = \sum_{c_1,c_2,v} A_{c_1 c_2 v}^{t} (\textbf{p}_{vc_1}\delta_{c c_2} - \textbf{p}_{vc_2}\delta_{c c_1}), 
\end{equation}
where $\textbf{p}_{vc}$ is the dipole matrix element between single-particle states $v$ and $c$. Transition energy $\Delta_{t}^{c}  ({\bf k}) =  E_{t}({\bf k}) - \varepsilon_{c}({\bf k })$ is a difference between the trion $E_{t}$ and single electron $\varepsilon_{c}$ energies, respectively. Normalization factor $Z$ is the Boltzmann statistical sum $Z= \sum_{\bf k}\sum_{t} \exp( - E_{t}({\bf k})/k_BT)$. Finally, in order to account for the finite lifetime of trion states we substitute the delta function in Eq. (\ref{eq:PL_spectrum}) by the Gaussian with width of $1$ meV.

\section{Results}

\subsection{PL spectrum} 

Before presenting results for the photoluminescence we calculate the linear absorption spectrum to compare with the earlier results. The calculations are done also using Eq. (\ref{eq:PL_spectrum}), where the temperature Boltzmann factor is omitted.

Figure~\ref{fig:abs} shows linear absorption spectra of MoS$_2$ monolayers, calculated for three doping densities. Obtained spectra are similar to the earlier results reported for MoS$_2$ monolayers.\cite{Qiu2013,Drppel2017,Zhumagulov2020}  For comparison Fig.~\ref{fig:abs} also shows the spectra of undoped MoS$_2$ monolayers, where peaks ``A'', ``B'' and ``A$_{2s}$'' are due to excitons. 

In doped monolayers, these exciton peaks split and the slitting grows with the doping. For peak ``A'' this can be interpreted as brightening of the intervalley exciton state.~\cite{Zhumagulov2020} For the sake of clarity, we refer to those peaks in doped monolayers simply as ``exciton peaks''  despite the three-particle nature of the corresponding states. In those states, an extra electron is bound to a two-particle exciton only loosely. In contrast, the lower ``T'' peak in Fig.~\ref{fig:abs} is due to the lower-energy trion states that have no counterpart in undoped monolayers. In these trion states both electrons are strongly localized near the hole.~\cite{Zhumagulov2020}

\begin{figure}[]
\begin{center}
\includegraphics[width=0.5\textwidth]{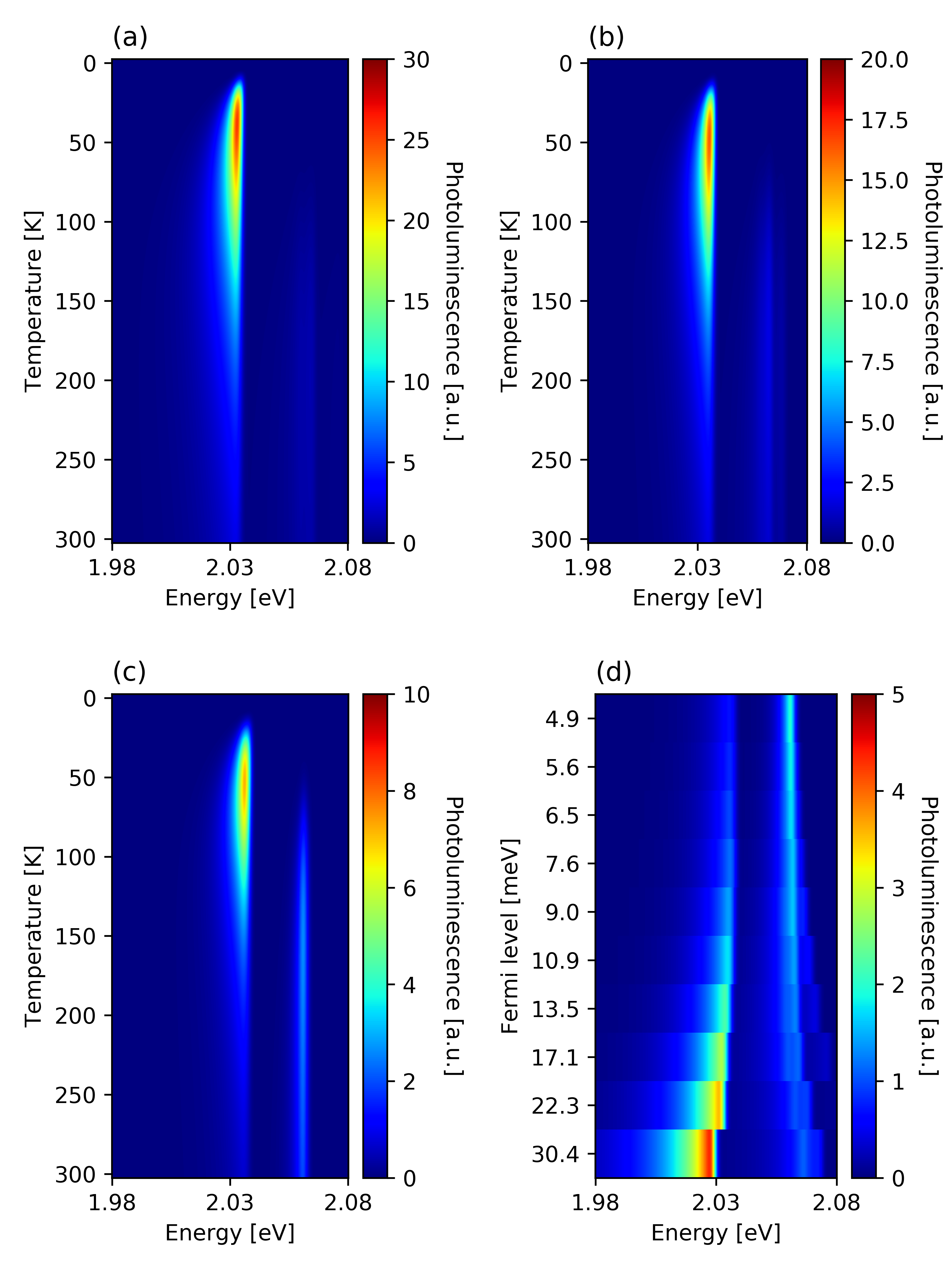}
\end{center}
\caption{Photoluminescence spectra of a MoS$_2$ monolayer.  Panels (a), (b) and (c) show the temperature dependence of the PL spectrum, calculated for (a) $n=19.7 \times 10^{11}$ cm$^{-2}$ ($E_F=17.1$ meV), (b) $n=12.58 \times 10^{11}$ cm$^{-2}$  ($E_F=10.9$ meV)  and (c) $n=5.6 \times 10^{11}$ cm$^{-2}$ ($E_F=4.9$ meV). Panel (d) shows the doping dependence (Fermi energy) of the PL spectrum, calculated at $T=300$ K.}
\label{fig:pl}
\end{figure}

Results of the calculations for the PL spectra are shown by colour density plots in Fig.~\ref{fig:pl}, where Figs.~\ref{fig:pl}a-c give the temperature dependence of the spectra, calculated for the same doping levels as the absorption spectra in Fig.~\ref{fig:abs}, and Fig.~\ref{fig:pl}d shows the doping dependence of the PL spectrum, obtained at $T=300$ K. As for the absorption spectra in Fig.~\ref{fig:abs}, results in Fig.~\ref{fig:pl} reveal two lowest energy peaks, trion and exciton, at $E\simeq 2.037$ eV and $E\simeq 2.060$ eV, respectively.

Comparing the temperature dependencies of the trion and exciton peaks in Fig.~\ref{fig:pl} one sees noticeable differences. The first one is that the exciton peak is practically absent at small temperatures, becoming visible only at $T\gtrsim 70$ K. This is easily understood by recalling that the exciton state is not occupied at small temperatures because of its larger energy. When the temperature raises the exciton occupation increases and so does the intensity of the corresponding spectral peak. After reaching its maximum, the peak intensity starts to decline but only modestly.  

Similarly, the lower trion peak in  Fig.~\ref{fig:pl} is practically absent at small temperatures, however, it becomes visible already at $T\gtrsim 10$ K. The peak intensity demonstrates a sharp non-monotonic $T$ dependence, rising quickly to its maximum at $T\simeq 65$ K before a fast drop at higher temperatures.

The doping dependencies of the trion and exciton peaks in Fig.~\ref{fig:pl}d demonstrate opposite trends. When the doping level increases, the trion peak shits to lower energies (redshift), whereas the exciton peak splits into two and moves to higher energies, in agreement with the earlier calculations.~\cite{Zhumagulov2020} One also notes that the intensity of the trion peak increases at large doping. The trion peak demonstrates another peculiar feature: it becomes asymmetric. At large temperatures, its width on the left side is notably larger than on the right side. Correspondingly, the total width of the peak also increases with $T$.

\begin{figure}[]
\begin{center}
\includegraphics[width=0.5\textwidth]{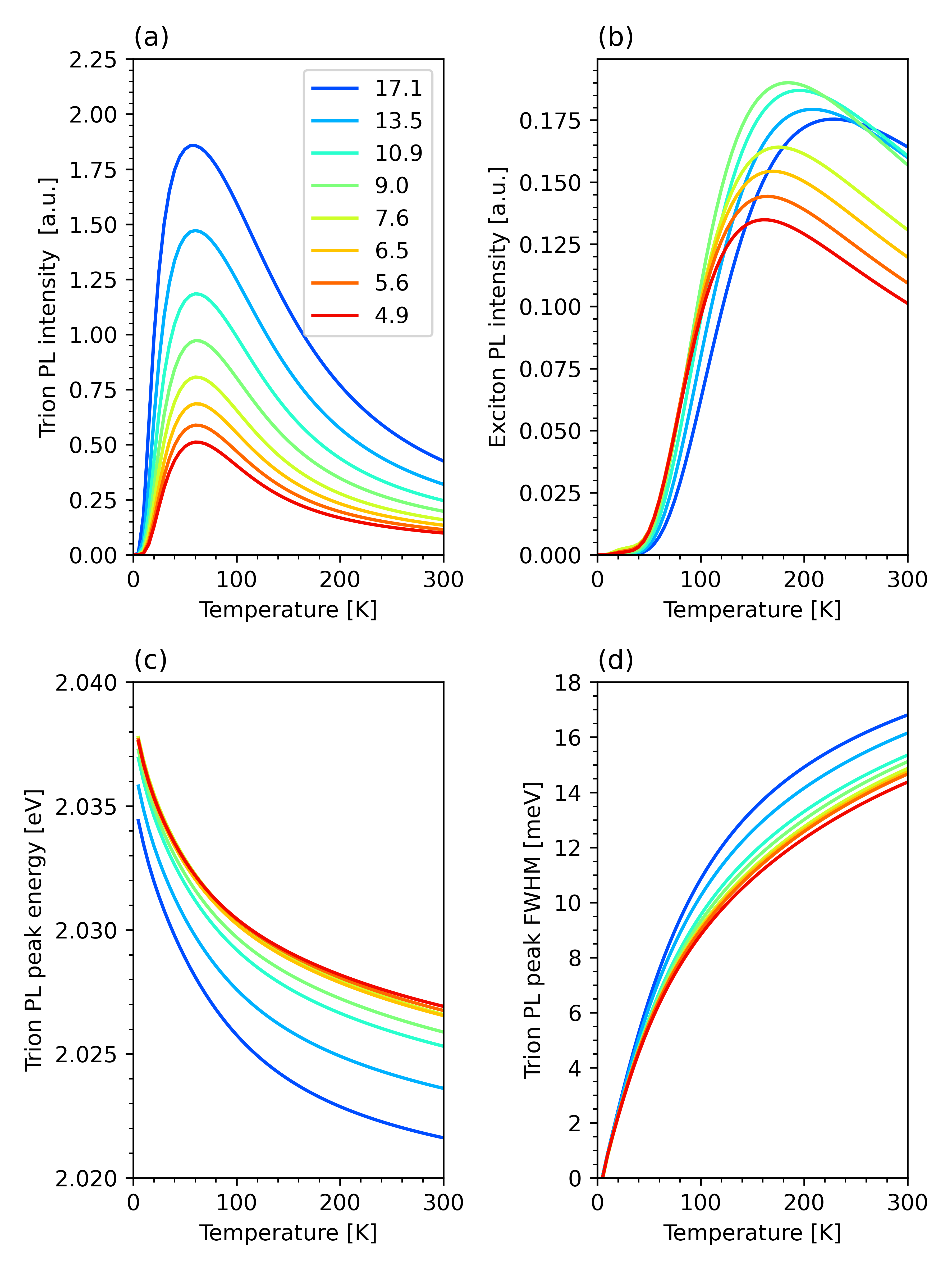}
\end{center}
\caption{Temperature dependence of the PL peak parameters: (a) intensity of the trion peak, (b) intensity of the exciton peak, (c) energy position of the trion peak and (d) width (FWHM) of the trion peak, calculated for several values of the Fermi energy (shown in the legend (a) in meV's).}
\label{fig:pl_res}
\end{figure}

Details of the temperature and the doping dependence of the trion peak are found in Fig.~\ref{fig:pl_res}. The temperature dependence of the trion peak intensity reveals a sharply non-monotonous behavior, discussed above, which is qualitatively similar for all considered doping values. However, the intensity of the trion peak grows notably when the doping increases. Interestingly, the position of the maximum of the temperature dependence is nearly independent of the doping.

The temperature dependence of the exciton peak in Fig.~\ref{fig:pl_res}b is also non-monotonic, but it is shifted towards the higher temperatures so that the maximum is reached at $T\simeq 200$ K. Its doping dependence is also non-monotonic: its intensity first increases with the doping and then start to decrease. However, in contrast to the trion peak, the position of the maximum now shifts with the doping notably.

The energy of the trion peak in the PL spectrum, shown in Fig.~\ref{fig:pl_res}c, reveals a monotonous downward trend when the temperature increases - the so-called redshift, which has a modest doping dependence.   Finally, the peak width, i.e. full width at half maximum (FWHM), in Fig.~\ref{fig:pl_res}d rapidly increases at larger temperatures. The broadening takes place in the absence of any temperature-dependent relaxation mechanisms such as phonon scattering not considered here. Here the broadening is induced by the trion states themselves and the matrix elements for the optical transitions.  The doping dependence of the broadening is also relatively weak but it increases at large temperatures.

\begin{figure}[]
\vspace{-0.cm}
\begin{center}
\includegraphics[width=0.5\textwidth]{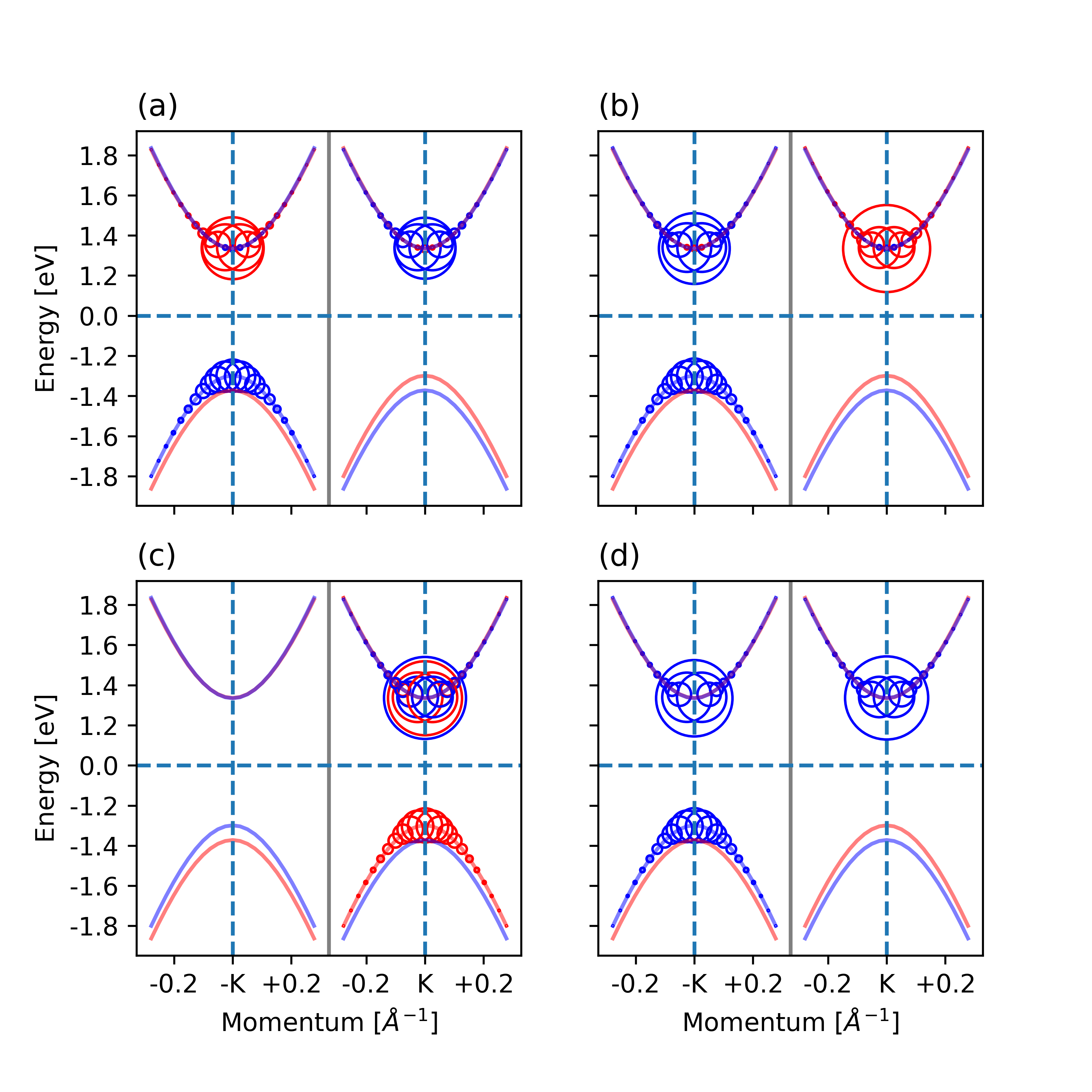}
\end{center}
\caption{Band structure of single-particle states and their contributions to the four lowest energy trion states, calculated for doping density $n=5.6 \times 10^{11}$ cm$^{-2}$. Panels (a), (b), (c) and (d) show trion states $T_D$, $T_1$,  $T_2$ and  $T_3$, respectively. A center of a circle indicates a contributing single-particle state, its radius denotes a relative weight in the trion state. Blue and red colours denotes the $z$-component of the particle spin $1/2$ and $-1/2$, respectively.}
\label{fig:wave}
\end{figure}

\subsection{Low energy trion states}

Observed characteristics of the ``T'' peak are related to properties of the lowest energy trion states. Our calculations reveal that there are four trion states with energies close to the position of the trion peak in the PL spectrum. Their characteristics are summarised in Table~\ref{tab:states}.

To illustrate the structure of the trion states, Fig.~\ref{fig:wave} shows the momentum dispersion of the single-particle states together with their relative contributions to the trion wavefunctions. Contributing single-particle states are marked by centers of the circles while their radii give weights of their relative contributions. The colors red and blue indicate the $z$ axis projection of the spin of the contributing states.

The ground state trion $T_D$ [Fig.~\ref{fig:wave}a] is dark because the recombination of electron and hole states in the same valley ($-K$) is not allowed due to the spin selection rules, whereas indirect recombination between electrons in the $K$ valley and holes in the $-K$ valley is strongly suppressed. In contrast, trion states $T_{1,3}$ [Fig.~\ref{fig:wave}b and Fig.~\ref{fig:wave}d] are bright because for these states the spin selection rule allows recombination of an electron and a hole in the same valley. Finally, the trion $T_{2}$ is also bright, however, this state comprises electrons and a hole in a single valley [Fig.~\ref{fig:wave}c].

We also note, that the energies of all bright trion states are very close. The energy difference among these states is on the scale of 1 meV, which makes it hard to distinguish them experimentally. The energy of the dark state is lower than that of the bright states by several meV.

\begin{center}
\begin{table}[]
    \centering
    \begin{tabular}{|c|c|c|c|c|c|}
        \hline
         trion  & type &$|\tau_{c_1} + \tau_{c_2}|$  & $|s^{z}_{c_1} + s^{z}_{c_2}|$   & $s^{z}_T\tau_T$ & $\Delta_{t}^c(0)$ [eV]\\ 
         \hline
         \hline
         $T_D$& dark  & 0 & 0 & $+1/2$&  2.033\\
         \hline
         $T_1$& bright  & 0 & 0 & $+1/2$&2.039\\
          \hline
         $T_2$& bright & 2 & 0 & $-1/2$& 2.037\\
          \hline
         $T_3$& bright & 0 & 1 & $-1/2$& 2.038\\
         \hline
    \end{tabular}
    \caption{Classification of the lowest energy trion states in a MoS$_2$ monolayer, calculated for the doping density $n=5.6 \times 10^{11}$ cm$^{-2}$.}
    \label{tab:states}
\end{table}
\end{center}

\subsection{Simple model for the PL spectra}

\begin{figure}[]
\begin{center}
\includegraphics[width=0.5\textwidth]{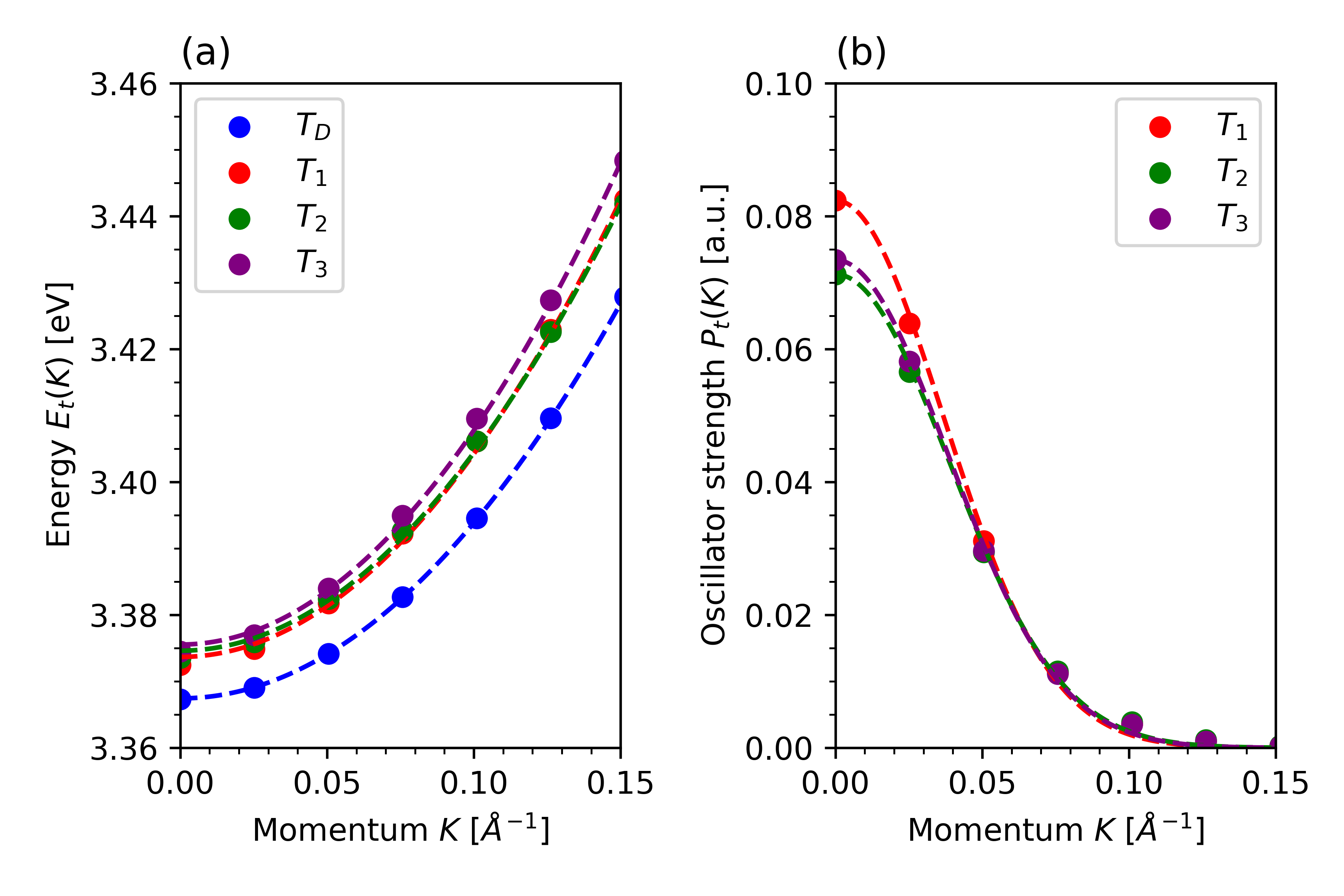}
\end{center}
\caption{Momentum dependence of the trion energy (a) and oscillation strength (b) of the bright trion states calculated for the doping density $n=5.6 \times 10^{11}$ cm$^{-2}$.
}
\label{fig:trion-momentum}
\end{figure}

Many properties of the trion peak in the PL spectra, in particular its temperature dependence, can be understood by adopting a simple phenomenological model, where trion excitations are regarded as quasi-particles with the energy dispersion approximated by the quadratic dependence at small momenta:  
\begin{equation}
    E_t({\bf k}) =  E_{t}(0)  + \frac{\hbar^2 {\bf k}^2}{2 M}.
    \label{eq:quadratic_dispersion}
\end{equation}
The validity of this approximation is illustrated in Fig.~\ref{fig:trion-momentum}a that plots the momentum dependence of the trion energy. The effective mass $M$ and the energy $E_{t}(0)$ are obtained by fitting the numerical results with Eq.~(\ref{eq:quadratic_dispersion}). Assuming the quadratic dispersion for single electrons as well, we write the optical transition energy as:
\begin{equation}
    \Delta_{t}^{c} ({\bf k}) =  \Delta_0  + \frac{\hbar^2 \textbf{k}^2}{2} \left( \frac{1}{M}-\frac{1}{m} \right).
     \label{eq:quadratic_difference}
\end{equation}
Taking into account that the effective electron mass in the Dirac spectrum is $m = 0.5$ m$_e$ and that $M\approx 3m > m$, one concludes that $\Delta_{t}^{c} ({\bf k})$ is a decreasing function of the trion momentum $k$.

In order to calculate the PL spectral function in Eq.~(\ref{eq:PL_spectrum}) we also need matrix elements corresponding to transitions between trions and delocalized electron states.  The latter are expressed as the form-factor of the trion wavefunction~\cite{Esser2000,Christopher2017}
\begin{equation}
    P_t^c(\textbf{k}) \propto \int d\textbf{r}^2 
    \Psi^t\left(\textbf{x}_1-\textbf{x}_3 = \textbf{r}, \textbf{x}_2=\textbf{x}_3 \right) e^{i \textbf{r} \textbf{k}}. 
\end{equation}
For a localized $\Psi^t$, the rates  $P_t^c$ can be well approximated by the Gaussian function\cite{Christopher2017}
\begin{equation}
    P_t^c (\textbf{k}) =  P_t (0) e^{- b^2 \textbf{k}^2},
    \label{eq:gaussian}
\end{equation}
where $b$ is the effective radius of the trion, which determines the spatial overlap between holes and  electrons in the trion wavefunction.  Numerical calculations of $P_t (\textbf{k})$, shown in Fig.  \ref{fig:trion-momentum}b, reveal that it is practically the same for all branches of bright trion states. 

We now calculate the PL spectrum by assuming that only these four trion states contribute to PL and, furthermore, that all bright trions have the same parameters.  Substituting Eqs. ~(\ref{eq:quadratic_difference}) and (\ref{eq:gaussian}) into Eq.~(\ref{eq:PL_spectrum}) and assuming for simplicity that the spectral lines are not broadened, we obtain the PL spectrum as:
\begin{align}
    L(\omega) &= \frac{P_t(0)}{k_BT}\,\frac{mM}{M-m} \, \frac{e^{ \alpha (\hbar\omega -  \Delta_0)} }{M_de^{\delta/k_BT} +3M}, 
    \label{eq:PL_estimate}
\end{align}
when $\hbar\omega < \Delta_0$, and it is zero, when the inequality is reversed. In this expression 
\begin{align}
    \alpha &=\frac{m}{M - m} \left( \frac{1}{k_BT} +  \frac{1}{k_BT_s}  \right)
\label{eq:alpha}
\end{align}
and $M_d$ and $M$ are effective masses of the dark and bright trions, $\delta = E_i(0) - E_D(0)$ is the difference between the energy of the bright and dark trions at $k=0$ and $k_BT_s=\hbar^2/(2Mb^2)$ is the characteristic trion temperature. This expression highlights the importance of the fact, that the ground state of the system is the dark trion. This fact gives rise to the exponential factor $\exp(\delta /k_B T)$ in Eq. (\ref{eq:PL_estimate}) that suppresses PL intensity at small temperatures.  

The energy dependence of the trion peak given by Eq. (\ref{eq:PL_estimate}) is asymmetric.  The origin of the asymmetry is the momentum dependence of the transition energy in Eq. (\ref{eq:quadratic_dispersion}) which is a decreasing function of the momentum. The width of the left wing of the spectral peak is $1/\alpha$ and it is temperature dependent as follows from Eq.~(\ref{eq:alpha}). The  total width of the peak (FWHM) is calculated as:
\begin{align}
    {\rm FWHM} \propto \frac{T}{1 + T/T_s},
    \label{eq:broadening}
\end{align}
which increases linearly at small $T$ and saturates when $T\gg T_s$, qualitatively reproducing numerical results in Fig.~\ref{fig:pl_res}d.  

The oscillator strength (OS) of the peak is obtained by integrating Eq.~(\ref{eq:PL_estimate}) over energy which yields: 
\begin{align}
   I_t \propto  \frac{1}{1 + T/T_s} \, \frac{1} {3+  \gamma e^{\delta/k_B T} },
    \label{eq:PLestimate}
\end{align}
where $\gamma = M_d/M$. This expression gives a non-monotonic temperature dependence for the OS, which rises at small $T$, reaches maximum at $T \sim \delta/k_B$ and then decreases at large temperatures, in agreement with the numerical results in Fig.~\ref{fig:pl_res}a. 

\begin{figure}[]
\begin{center}
\includegraphics[width=0.5\textwidth]{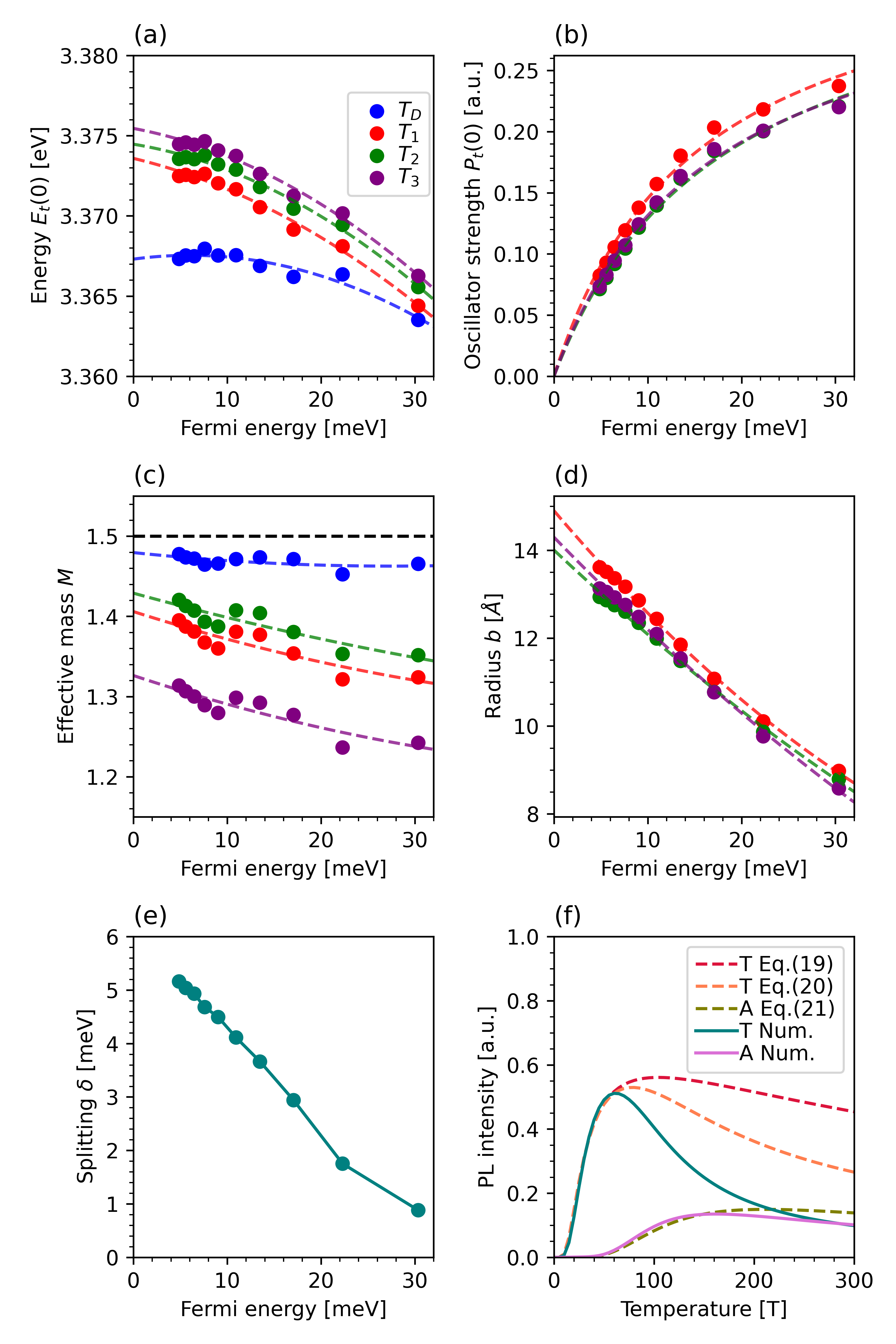}
\end{center}
\caption{Doping dependence of parameters of the trion states: (a) zero-momentum energy $E_t(0)$, (b) zero momentum OS $P_t(0)$, (c) effective mass $M$, (d) effective radius $b$ and (e) energy splitting $\delta$ between the dark and bright trions. Black dashed line in panel (c) gives a bare trion mass - a sum of the masses of two electrons and one hole (1.5 $m_e$). Panel (e) compares numerical results for the OS of the trion and exciton peaks calculated using Eqs.~(\ref{eq:PLestimate}), (\ref{eq:PLestimate2}) and (\ref{eq:PLexc}). } 
\label{fig:doping}
\end{figure}

For a more quantitative comparison with the numerical results it is important to note that quantities that enter Eq. (\ref{eq:PL_estimate}) depend on the doping. In most cases this dependence cannot be described with a simple model, with the exception of $ P_t (0) \propto  E_F/(E_0+E_F)$, where $E_0 \simeq 16.8$meV for bright trions. The doping dependence of the quantities entering Eq.~(\ref{eq:PL_estimate}), extracted from the numerical calculations, is plotted in Fig.~\ref{fig:doping}, which gives the energy  $E_t(0)$ of trions at $k=0$ [Fig.~\ref{fig:doping}a], the OS $P_t(0)$ of trion states at $k=0$ [Fig.~\ref{fig:doping}b],  the effective mass $M$ [Fig.~\ref{fig:doping}c] and the radius $b$ of trions [Fig.~\ref{fig:doping}d], as well as the  splitting energy $\delta$ between the dark and bright trions [Fig.~\ref{fig:doping}e]. 

To compare the OS estimate, given by Eq.~(\ref{eq:PLestimate}), with the numerical results we take the doping density $n= 5.6\times 10^{11}$cm$^{-2}$ and use values for the parameters given in Fig.~\ref{fig:doping}. The comparison is shown in  Fig. \ref{fig:doping}f, revealing an excellent quantitative agreement between the estimate and the numerical results, but only at $T \lesssim 65$ K. At higher temperatures Eq.~(\ref{eq:PLestimate}) overestimates the OS considerably. 

The mismatch is explained by the fact that Eq.~(\ref{eq:PLestimate}) does not take into account higher energy states, that are also occupied at larger temperatures. In order to demonstrate the influence of these states, we calculate the OS using the same simple model but taking into account the higher energy states that correspond to the exciton peak. This modifies Eq. (\ref{eq:PLestimate}) for the OS of the lowest trion peak as:  
\begin{align}
    \label{eq:PLestimate2}
   I_{t}^\prime &\propto  \frac{1}{1 + T/T_s}\, \frac{1} {3+ \gamma  e^{\delta/k_B T} + g_e e^{-\delta_{e}/k_BT }}
\end{align}
where $g_e=8$ and $\delta_{e}$ is the energy difference between exciton and bright trion states.~\cite{Zhumagulov2020} The corrected OS is also plotted in Fig.~\ref{fig:doping}f, demonstrating a better agreement with the numerical results at $T\gtrsim 65$ K (we use $\delta_e = 23$ meV extracted from the numerical calculations [Fig.~\ref{fig:abs}]). Accounting for states with the larger energy would further improve this agreement. 

The same conclusion holds for the OS of the exciton peak. In this case one needs to account for the states with the energy less than that of the peak. This yields the OS in the form
\begin{align}
   I_{e} &\propto  \frac{1}{T}\, \frac{e^{-\delta_{e}/k_BT }} {3+ \gamma  e^{\delta/k_B T} +g_e e^{-\delta_{e}/k_BT }}.
\label{eq:PLexc}
\end{align}
This estimate is also plotted in Fig.~\ref{fig:doping}f alongside the numerical results for the exciton peak OS [cf Fig.~\ref{fig:pl_res}b], demonstrating a fairly good agreement in the full range of temperatures considered here. 

Comparing results for the trion peak intensity in Fig.~\ref{fig:pl_res}a and the doping dependence of the parameters in Eq.~(\ref{eq:PL_estimate}), given in Fig.~\ref{fig:doping}, one notices an interesting fact: while the trion splitting energy $\delta$ demonstrates a very strong doping dependence [Fig.~\ref{fig:doping}e], the temperature of the maximal intensity $T_{\rm max} \simeq 65$ K  is almost independent of the doping [Fig.~\ref{fig:pl_res}a]. It is explained by that the doping dependencies of the parameters $\gamma$ and $T_s\propto M^{-1}b^{-2}$ in Eq.~(\ref{eq:PLestimate}) compensate that of $\delta$. Thus the estimate $T_{\rm max}  \simeq \delta/k_B$, while qualitatively correct, can be quite far quantitatively.   

Finally, we mention that the same approach can be used to estimate the temperature dependence for the redshift of the trion peak. In order to derive it, one calculates the PL spectrum  using Eq. (\ref{eq:PL_spectrum}) where the delta function is substituted by the bell shape with a finite width (we do this also in the numerical calculations). In this case the calculations yield a more complex expression for the spectrum. However, one can obtain a simple estimation of this result according to which the redshift of the trion peak is given as $\Delta E \simeq  - {\rm FWHM}$, where FWHM is defined in Eq.~(\ref{eq:broadening}).  This estimate yields that the absolute value of the shift $\Delta E$ grows linearly with temperature at small $T$ but saturates when $T$ increases, in full agreement with the numerical result in Fig.~\ref{fig:pl_res}c.   
 
\section{Conclusions}

We investigate the temperature and doping dependencies of the lowest energy trion peak in the PL spectra of a doped MoS$_2$ monolayer. The calculations are done using a combined approach, where single-particle states are described using the effective Dirac Hamiltonian, while the trion states are calculated by a direct diagonalization of the three-particle Hamiltonian within the Tamm-Dancoff approximation. 
The PL spectrum is calculated under the assumption that excited states reach thermal equilibrium on timescales much shorter than the radiative recombination time of excitons and trions. 

Our calculations demonstrated that the lowest energy peak of the PL spectrum is determined by four trion states: three bright and one dark. The temperature dependence of the trion PL spectral peak demonstrates features,  markedly different from that of the exciton peaks. The temperature dependence of the trion peak intensity is sharply non-monotonic with a maximum at $T\simeq 65$ K, in contrast with the lowest energy exciton peak with the maximum at  $T \simeq 200$ K. The trion peak also exhibits a strong redshift and an asymmetric broadening even in the absence of the electron-phonon interaction.

Using a simple model, that takes the momentum dependence of the energy and transition matrix elements into account, we can explain the temperature dependencies of the PL spectra and to provide simple expressions for all pertinent characteristics of the trion peak. We expect similar conclusions to hold for other 2D transition metal dichalcogenides with valley degeneracy, which are awaiting further experimental verifications.


{\it Acknowledgements}
The work was supported by the Russian Science Foundation under the grant 18-12-00429. We also acknowledge support from the Deutsche Forschungsgemeinschaft (DFG, German Research Foundation – Project-ID 314695032 – SFB 1277) and the Alexander von Humboldt Foundation and Capes (Grant No. 99999.000420/2016-06).

\bibliography{bibliography}

\end{document}